\documentclass[draft, ms]{agutex2015}

\usepackage{lineno}
\usepackage{graphicx}
\usepackage{amsmath}
\usepackage{amssymb}
\usepackage{xcolor}

\setkeys{Gin}{draft=false}

\authorrunninghead{WANG ET AL.}

\titlerunninghead{ENKF-BASED TSUNAMI INVERSION}

\authoraddr{Corresponding author: Heng Xiao,
Department of Aerospace and Ocean Engineering, Virginia Tech, Blacksburg, VA 24060, USA. Email:
hengxiao@vt.edu; Tel: +1 540 231 0926}

\begin{document}

%
%

\title{Inversion of Tsunamis Characteristics from Sediment Deposits Based on Ensemble Kalman
  Filtering}



\authors{Jian-Xun Wang,\altaffilmark{1}
Hui Tang,\altaffilmark{2} Heng Xiao,\altaffilmark{1}
Robert Weiss\altaffilmark{2}}

\altaffiltext{1}{Department of Aerospace and Ocean Engineering,
Virginia Tech, Blacksburg, Virginia, USA.}

\altaffiltext{2}{Department of Geosciences,
Virginia Tech, Blacksburg, Virginia, USA.}

%
%

 \keypoints
 {
 \item Proposed a rigorous, Bayesian framework for tsunami inversion based on sediment deposits.
 \item Used Ensemble Kalman filtering for inversion, providing quantified uncertainties in the results.
 \item Verified and demonstrated inversion performance based on realistic setting of sediment deposits
 }

%
%

\begin{abstract}
  Sediment deposits are the only records of paleo tsunami events.  Therefore, inverse
  modeling methods based on the information contained in the deposit are indispensable for
  deciphering the quantitative characteristics of the tsunamis, e.g., the flow speed and the flow
  depth. In this work, we propose an inversion scheme based on Ensemble Kalman Filtering (EnKF) to infer tsunami
  characteristics from sediment deposits. In contrast to traditional data assimilation methods using
  EnKF, a novelty of the current work is that we augment the system state to include both the
  physical variables (sediment fluxes) that are observable and the unknown parameters (flow speed
  and flow depth) to be inferred.  Based on the rigorous Bayesian inference theory, the inversion
  scheme provides quantified uncertainties on the inferred quantities, which clearly distinguishes
  the present method with existing schemes for tsunami inversion. Two test cases with synthetic
  observation data are used to verify the proposed inversion scheme. 
  Numerical results show that the tsunami characteristics inferred from the sediment deposit information
  agree with the synthetic truths very well, which clearly demonstrated the merits of the 
  proposed tsunami inversion scheme. {Furthermore, a realistic application of the proposed inversion
  scheme with the field data from the 2006 South Java Tsunami is studied, and the results are compared to the
  previous inversion model results in the literature and are validated with the field data. The 
  comparisons show excellent performance of the proposed inversion scheme in realistic applications.}
\end{abstract}

%
%

\begin{article}

\section{Introduction}
\label{sec:intro}
In recent decades, coastal cities have become important nodes in global economic network. Therefore,
adverse impacts from coastal disasters, such as tsunamis, do not only have local or regional
effects, but can amount to global consequences. Furthermore, the increased economic relevance of
cities, such as Los Angeles, Singapore or Hong Kong (to mention only three of many), has caused
their population to significantly grow, including projections that even more people will be
attracted to these economic power houses.  Because the population of coastal megacities increased
and is predicted to grow even more, the risk from coastal disasters to which the coastal population
is exposed needs to be carefully, realistically and objectively evaluated. To create risk
assessments that meet these attributes, meaningful and robust hazard assessments are
required. Fortunately for coastal megacites, not enough tsunamis have occurred in any one region to
solely base hazard assessments on historical or modern record events. Because of that fact, the geologic
record of paleo tsunamis needs to be interrogated. The important nature of the geologic record is
the presence of uncertainty. This is not only because of the chaotic behavior of tsunami sources,
such as earthquakes, landslide, volcanic eruptions, and meteorite impacts, but also due to the fact
that, for example, storms can produce deposits with features similar to those preserved in tsunami
deposits.  There also are other sources of uncertainty.  For example, the fact that the
sedimentation process depends very strongly on the presence of local turbulence is an additional
source of uncertainty that can only be reduced by a better understanding of sediment transport
processes.  The presence of uncertainty is one of the arguments employed to base tsunami hazard
assessment on statistics.

To include the information contained in tsunami deposits, the often qualitative information
retrieved from deposits needs to be translated into quantitative data about the causative event.
Tsunami inversion models have been designed to estimate the flow conditions at the time of
depositions. Models such as the Moore's Advection Model ~\citep{Moore2007336}, Soulsby's Model
~\citep{sousby2007} and TsuSedMod ~\citep{jaffe2007simple} were develop to retrieve quantitative
data from the deposits about the causative event. While these models have been successfully
applied~\citep{Moore2007336,sousby2007,jaffe2007simple,Spiske201029,Jaffe201123,Jaffe201290,witter2012,Spiske201331},
the problem, however, is that these models cannot be seen as inversion models by a more strict
definition of inversion. Specifically, important features of inversion models, such as uncertainty
quantification and error analysis, cannot be properly carried out. To clarify, an inversion model
performs an inversion based on a forward model and an inversion scheme. Without playing down the
important impact the above-mentioned models have had, the present methods are more data-fitting
models than inversion models in a strict, mathematical sense.

In this contribution, we propose a rigorous, Bayesian scheme for tsunami inversion based on Ensemble
Kalman Filtering (EnKF). Data assimilation methods based on EnKF are widely used in geosciences
applications such as numerical weather forecasting, where the initial conditions (present states of
the system, e.g., pressure, velocity, and temperature) are not known precisely and thus must be
inferred from observations.  The novelty of the present work is the augmentation of the system state
to include the unknown parameters and the use of EnKF-based data assimilation method to infer these
parameters.  To the authors' knowledge, our work represents the first attempt to use EnKF for
tsunami inversion.  While this contribution should be seen as a proof of concept, a more
comprehensive parameter study has been performed and will be presented in a separate paper.  The
proposed method would open up possibilities for mathematically more rigorous tsunami inversion from
deposits with quantified uncertainties.

\section{Methodology}
\label{sec:method}
The objective of this work is to infer tsunami flow characteristics from the tsunami deposits, which
is an inverse problem. As with most inversion algorithm, the proposed method for the inverse problem
involves solving the forward problem repeatedly, i.e., computing the tsunami deposit from given
tsunami flow characteristics. Therefore, in this section the formulations, assumptions, and solution
algorithm of the forward problem is first presented in Section~\ref{sec:met-for}. Subsequently, the
formulation and solution algorithm for the inverse problem is introduced in
Section~\ref{sec:met-inv}.

\subsection{Formulation of the Forward Problem and Solution Algorithm}
\label{sec:met-for}

\begin{figure}[!htbp]
  \centering
  \noindent
  \includegraphics[width=0.9\textwidth]{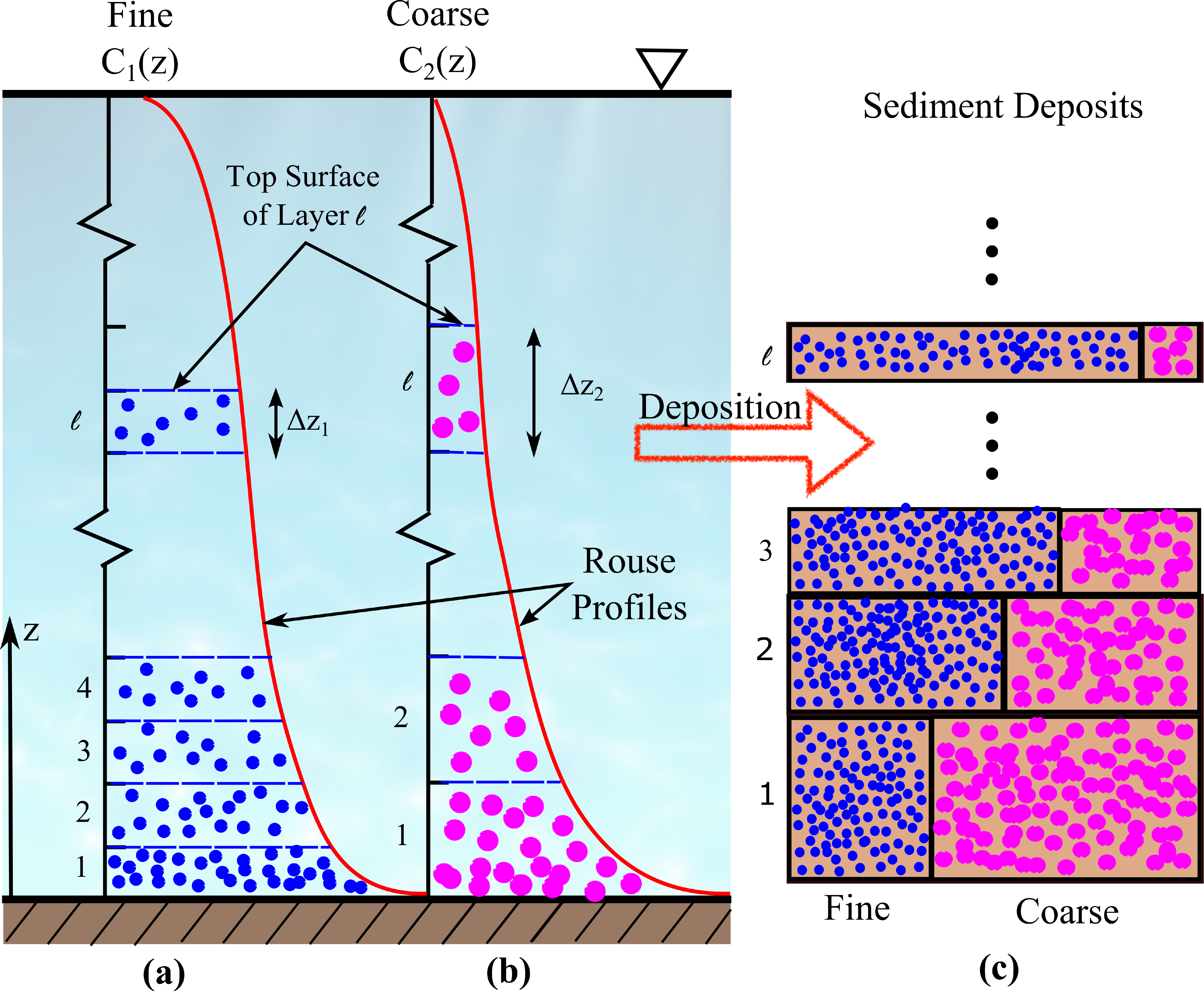}
  \caption{A schematic illustration of the algorithm for solving the forward problem, i.e.,
    computing the tsunami deposit thickness and the grain-size distribution in each layer from
    given tsunami characteristics (flow speed and flow depth). An exemplary output of the algorithm
    is presented in Fig.~\ref{fig:sed}. }
  \label{fig:forward}
\end{figure}

Given the tsunami characteristics (i.e., flow speed $u$ and flow depth $h$), the forward problem is
to compute the sediment deposition characteristics (i.e., the depth of the tsunami deposit and the
particle size composition at any height of the sediment column). An example output of the forward
problem is shown in Fig.~\ref{fig:forward}c, which shows the thickness of the tsunami deposit, the
grain-size distribution in each layer.

The forward model adopted in this study is based on the simplified sedimentation models
of~\citet{jaffe2007simple} and~\citet{huitang2015}. In these models it is assumed that the tsunami
deposit is formed solely by the steady sedimentation of the particles in the water column. The
effects of bed load transport and the acceleration of sediment particles during the settling process
are neglected.  It is further assumed that the sediment concentration and the fluid velocity vary
only vertically in the water column, and their horizontal gradients and temporal changes are
neglected.  With the assumptions above, the flow velocity profile $u(z)$ along the water column can
be parameterized by the shear velocity $u_*$, where $z$ is the elevation above the bed.
Consequently, the depth-averaged flow velocity can be obtained from the following integral over the
entire water column:
\begin{equation}
U = \int_{z_0}^{h}\frac{u_*^2}{K(z)} dz , 
\label{eq:uz}
\end{equation}
in which $z_0$ is the total roughness of the bed.  The eddy viscosity $K$ can be calculated
as following~\citep{gelfenbaum1986experimental}:
\begin{equation}
\label{eq:eddy}
K(z) = \kappa \, u_* \, z \exp\left[ {\frac{-z}{h} - 3.2 \left( \frac{z}{h} \right)^2 + \frac{2}{3} \times 3.2
    \left( \frac{z}{h} \right)^3} \right],
\end{equation}
where $\kappa =0.41$ is the von Karman constant. 
The suspended sediment concentration for each grain size in the water column is assumed to follow the
Rouse profile~\citep{jaffe2007simple}:
\begin{equation}
C_i (z) = C_{i, 0} \; \exp\left[ {w_{i} \int_{z_0}^{z} {\frac{1}{K(z)}}} \right], 
\label{eq:ci}
\end{equation} 
in which $w_{i}$ is the settling velocity of particles in the $i^{\mathrm{th}}$ grain-size class,
which depends on the mean particle diameter $D_i$ of the class; $C_{i, 0}$ denotes sediment
concentration of the $i$\textsuperscript{th} grain-size class at the bed, which depends on the shear
velocity $u_*$ and the fraction of bed sediment in class $i$, among other
parameters~\citep{madsen1993wind}.  Based on Eqs.~(\ref{eq:uz}) and~(\ref{eq:ci}), the velocity
and concentration profiles can be uniquely parameterized by two scalar quantities, the shear
velocity $u_*$ and flow depth $h$. Therefore, from here on we consider $u_*$ and $h$ the
characteristics that define a tsunami. The objective of a tsunami inversion is thus to infer $u_*$
and $h$ from a tsunami deposit record.

Equation~(\ref{eq:ci}) suggests that the sediment concentration for each sediment grain size can be
determined by the tsunami characteristics (i.e., shear velocity $u_*$ and flow depth $h$) and the
particle diameter $D_i$. According to the convention in the sedimentology, the grain size is
represented in $\phi$ scale (the logarithm of the particle diameter to the base 2), $\phi_i =
-\log_2{(D_i/D_{\textrm{ref}})}$, in which $D_{\textrm{ref}} = 1$~mm is the reference particle
diameter to ensure dimensional consistency.  Consequently, the tsunami deposit thickness and
grain-size distribution can be obtained by integrating the concentration curve $C_i(z)$ for each
grain size class.  The integration is performed with the following algorithm, which is illustrated
in Fig.~\ref{fig:forward} with two grain-size classes as an example.

\begin{enumerate}
\item \textbf{Discretization of time.} Assuming that $T$ is the total time taken for all the
  sediment in the water column (including all grain-size classes) to settle, we divide the time $T$
  to $N$ time steps of size $\Delta t$ such that $T = N \, \Delta t$.

\item \textbf{Discretization of water column.} For grain-size class $i$, the water column can be
  divided to $N$ layers, numbered sequentially upward from the bottom (see Fig.~\ref{fig:forward}a),
  such that the sediment at the top of layer $l$ arrives at the bed at time $l \, \Delta t$.  Since
  the particles are assumed to have no acceleration (e.g., at constant velocity $w_i$) during the
  sedimentation, the layers of the water column have a uniform thickness of $\Delta z_i = w_i \,
  \Delta t$.  Note that the layer thickness can be different among different grain-size classes
  since the terminal velocity $w_i$ is larger for coarse grains than for fine grains. Consequently,
  for a coarse grain-size class the discretized water column layer thickness $\Delta z_i = w_i
  \Delta t$ is larger and thus the number of discretized water column layers is smaller. This can be
  seen by comparing Fig.~\ref{fig:forward}a and~\ref{fig:forward}b.

\item \textbf{Accumulation of tsunami deposit.} With the discretization of the water column, it can
  be seen that the obtained tsunami deposit has $N$ layers as well (numbered in the same way as for
  the water column; see Fig.~\ref{fig:forward}c). The sediment in layer $l$ consists of the sediment
  in the $l$\textsuperscript{th} layer of the water column for all grain-size classes (indicted in
  colors/patterns in Fig.~\ref{fig:forward}c). The tsunami deposit thickness $\Delta \eta_l$ of the
  $l$\textsuperscript{th} layer is computed by summing up the sediment volume in the corresponding
  $l$\textsuperscript{th} water column layers for all grain-size classes:
  \begin{equation}
    \label{eq:d-eta}
    \Delta \eta_l = \frac{1}{C_0} \left( \sum_{i = 1}^{n} {\overline{C}_{i, l} \, \Delta z_{i}}
    \right), 
  \end{equation}
  in which $C_0$ is the total sediment concentration at the bed including size classes, $n$ is the
  number of grain-size classes, and $\overline{C}_{i, l}$ is the average concentration of grain-size
  class $i$ in the water column layer $l$, which can be obtained by a simple integration:
  \begin{equation}
    \label{eq:cbar}
    \bar{C}_{i, l} = \frac{1}{\Delta z_i} \int_z^{z + \Delta z_i} \, C_i (z) \, dz.
  \end{equation}

\item \textbf{Post-processing for grain-size distribution.}  The fraction $f_{i, l}$ of each
  grain-size class $i$ in the $l$\textsuperscript{th} layer in the sediment column is
  \begin{equation}
    \label{eq:frac}
    f_{i, l} = \frac{1}{C_0} \frac{\overline{C}_{i, l} \, \Delta z_{i, l}}{\Delta \eta_l}.
  \end{equation}
\end{enumerate}
The thickness of sediment layers $\Delta \eta_l$ and the fraction $f_{i, l}$ for each grain-size
class in each layer are used to produce the plots presented in Fig.~\ref{fig:forward}c and
Fig.~\ref{fig:sed}, which is the final output of the forward problem.

The algorithm above for the forward problem produce some key information, which is
the time stamp of tsunami deposit layers and the sediment flux at each time step.  Specifically,
the time at which the sediment layer $l$ finished the deposition at time $l \, \Delta t$, which can
be considered the time stamp on the sediment layer (see Fig.~\ref{fig:forward}b).  Acknowledgedly,
the sediment cores obtained in the field \emph{do not} come with time stamps.  However, by comparing
sediment cores obtained at several locations with cross-shore offsets, it is possible to infer the
time stamps and thus the sediment fluxes at discretized time intervals~\citep{DEARING1981356}.

To facilitate the filtering procedure to be used in the inversion, we adopt an alternative approach
of computing tsunami deposit thickness by considering the time sequence of the sedimentation
process. Specifically, in contrast to Eq.~(\ref{eq:d-eta}), the deposition thickness can be computed
from the average sediment flux $\zeta_{i, l}$ of grain-size class $i$ at time step $l$ when the
$l$\textsuperscript{th} layer of the sediment deposit is formed. That is,
\begin{subequations}
    \label{eq:d-eta-flux}
    \begin{align}
    \label{eq:d-eta2}
    \Delta \eta_l & = \frac{1}{C_0} \left( \sum_{i = 1}^{n} \zeta_{i, l} \, \Delta t      
    \right),  \quad \textrm{with} \\
\begin{split}
  \label{eq:flux}
  \zeta_{i, l} & = \frac{\overline{C}_{i, l} \, \Delta z_{i}}{\Delta t},   \quad \textrm{or
    equivalently,} \\
  \zeta_{i, l} & =  \overline{C}_{i, l} \;  w_i  \quad \textrm{ in which }  \quad
  w_i = \frac{\Delta z_{i}}{\Delta t} .
\end{split}
\end{align}
\end{subequations}
Note that the same subscript $l$ that is used above as the indices of the water column layer (
Fig.~\ref{fig:forward}a and~\ref{fig:forward}b) and sediment layer (Fig.~\ref{fig:forward}c) is
used to denote the time step index here. This choice of notation is justified by the assumption that
the $l$\textsuperscript{th} sediment layer is formed by the deposition of all sediment grain classes
in the $l$\textsuperscript{th} water column layer at time step $l$.  Simply substituting
Eq.~(\ref{eq:flux}) to Eq.~(\ref{eq:d-eta2}) yields Eq.~(\ref{eq:d-eta}), and thus the two
formulations in Eqs.~(\ref{eq:d-eta}) and~(\ref{eq:d-eta-flux}) are equivalent.

By adopting the flux-based formulation, we are modeling the sediment deposition process as the
evolution of a dynamical system with the sediment flux $\zeta_i$ for each grain-size class $i$ as
the system state.  Based on the assumptions from Eq.~\ref{eq:flux}, it can be seen that for any
grain-size class $i$ the state $\zeta_i (t)$ is uniquely determined by the concentration profile
$C_i (z)$ and the settling velocity $w_i$, which in turn depends on the tsunami characteristics
(shear velocity $u_*$ and flow depth $h$) and the grain-size $\phi_i$.  Therefore, the forward
model $\mathcal{F}$ is thus formulated as to compute sediment deposition flux $\zeta_i(t)$ from
known shear velocity $u_*$ and flow depth $h$, i.e., $\mathcal{F}: (u_*, h) \mapsto \zeta_{i}
(t)$. The sediment thickness and grain-size distribution are considered auxiliary quantities
obtained by post-processing the time series of the system state $\zeta_{i} (t)$.

\subsection{Formulation of the Inverse Problem and Solution Algorithm}
\label{sec:met-inv}

The objective of the inverse problem is to infer the tsunami characteristics (depth-averaged flow
velocity and flow depth) from tsunami deposit.  In the formulation of the forward problem above,
the flow velocity profile and the depth-averaged velocity are parameterized by the shear velocity as
in Eq.~(\ref{eq:uz}), and the sediment flux is the state variable of the system. Therefore, the
inverse problem is recast as inferring the shear velocity and flow depth from the sediment flux. The
sediment flux, which is the input to the inverse problem, can be obtained by analyzing the tsunami
deposit from the field.  Specifically, when tsunami deposit samples are cored, they are first
divided to a number of layers and to obtain the time stamp for each layer by utilizing the spatial
information of the samples.  Particle-size analysis is then performed on each layer, i.e., by using
sieve analysis or other sedimentation techniques~\citep{martin1986}, which leads to the average
sediment flux $\zeta_{i}$ at a few discrete time steps. The inverse problem needs to be formulated
so that shear velocity $u_*$ and flow depth $h$ can be inferred from the sediment flux. As such, we
consider $u_*$ and $h$ the unobservable parameters of the dynamical system $\mathcal{F}(u_*, h;
\zeta_i (t))$.  They will be inferred from observations of the system state, i.e., the sediment flux
$\zeta_i (t)$.

The inversion of shear velocity and flow depth from observed sediment flux is  challenging
for at least two reasons. First, the observation is inevitably sparse and noisy, because the
sediment core can only be divided to a few layers to ensure each layer has enough sediment mass, and
the measurement has large errors. Second, the forward model describing the sedimentation process is
based on high simplified assumptions and thus does not faithfully represent the exact system
dynamics. 

In this work we use the Ensemble Kalman Filtering (EnKF) to perform the
inversion~\citep{evensen2003ensemble,evensen2007data,iglesias2013ensemble}, which is widely used in
data assimilations, particularly in numerical weather forecasting. When used to solve the tsunami
deposit inversion problem, the system state is first augmented to include both the physical state
$\zeta_i (t)$, which are observable, and parameters $u_*$ and $h$, which are unobservable (from the
sediment core) and are to be inferred. The augmented system state $\mathbf{x}(t)$ is written as a
vector formed by stacking the unknown parameters and the sediment flux $\zeta_i(t)$:
\begin{equation}
  \label{eq:aug-state}
  \mathbf{x} = [\zeta_1, \cdots, \zeta_n,  u_*, h]',
\end{equation}
in which $'$ indicates vector transpose.

Given the prior distributions for parameters ($u_*$ and $h$) to be inferred and the covariance
matrix $\mathbf{R}$ of the sediment flux observations $\zeta_{i}^{obs}$, the inversion algorithm
proceeds as follows:
\begin{enumerate}
\item \textbf{Sampling of prior distribution.} From the prior distributions of the parameters, $M$
  samples are drawn. Each sample consists of a combination of values for $u_*$ and~$h$.
\item \textbf{Propagation.}  The sediment fluxes $\hat{\zeta}_{i}$ for all grain-size classes are
  computed from Eq.~\ref{eq:flux} by using the updated parameters $u_*$ and $h$ from the previous
  analysis step (or from the initial sampling if this is the first propagation step). The
  propagation is performed for $\Delta N$ time steps, where $\Delta N$ is the number of time steps
  in the time interval $\Delta T$ between two consecutive data assimilation operations.  The
  $\hat{\cdot}$ indicates predicted quantities that will be corrected in the analysis step below.
  The propagation is performed for each sample in the ensemble, leading to the propagated ensemble
  $\{\hat{\mathbf{x}}_j\}_{j = 1}^M$.  Each sample $\hat{\mathbf{x}}_j$ is a vector containing a
  realization of shear velocity and flow depth as well as the corresponding sediment flux (see
  Eq.~(\ref{eq:aug-state})).  The mean $\bar{\mathbf{x}}$ and covariance $\mathbf{P}$ of the
  propagated ensemble are computed (see Eq.~(\ref{eq:cov}) in Appendix~\ref{app:enkf}).

\item \textbf{Analysis/Correction.}  The computed sediment fluxes $\hat{\zeta}_{i}$ for all
  grain-size classes are compared with observations $\zeta_{i}^{obs}$ corresponding to the current
  time step $l$.  The ensemble covariance $\mathbf{P}$ and the error covariance $\mathbf{R}$ are
  used to compute the Kalman gain matrix $\mathbf{K}$.  Each sample is corrected as follows:
\begin{equation} 
  \label{eq:analysis}
  \mathbf{x}_j = \hat{\mathbf{x}}_j + \mathbf{K} ({\boldsymbol{\zeta}_j} - \mathbf{H}
  \hat{\mathbf{x}}_j)  
\end{equation} 
where superscript $\mathbf{x}_j$ is the corrected system state; $\boldsymbol{\zeta} = [\zeta_1,
\cdots, \zeta_n]'$ are the sediment fluxes, the part of the system state vector that can be
observed; $\mathbf{H}$ is the observation matrix. After the correction, the analyzed state contains
updated fluxes and parameters.

It should be remarked that: (1) the analysis scheme above suggest that the corrected state (i.e., the analysis)
is a linear combination of the prediction and observations, with the Kalman gain matrix $\mathbf{K}$
being the weight of the observations; (2) the observation matrix $\mathbf{H}: \mathbb{R}^{n+2}
\mapsto \mathbb{R}^n$ has a size of $n \times (n+2)$, which maps a vector in the $n+2$ dimensional
state space to a vector in the $n$ dimensional observation space.  The first $n$ columns of
$\mathbf{H}$ are ones and the last two columns are zeros, indicating that the parameters are not
observed.

\item Repeat propagation and analysis steps 2--3 for next data assimilation time $t+\Delta T$ to
  incorporate next observation until all observations are assimilated.  The EnKF algorithm for
  inferring tsunami characteristics from tsunami deposit is summarized in
  Fig.~\ref{fig:enkfFlow}. The detailed algorithm is presented in Appendix~\ref{app:enkf}.
\end{enumerate}

 It can be seen that the observations arrive sequentially in the EnKF data
assimilation procedure above, which is typical for applications such as numerical weather
forecasting. In this work we formulate the tsunami inversion problem with a sequential streaming of
data to take advantage of the widely used EnKF algorithm. In this method, the filtering procedure
finds an optimal correction at each assimilation step based on the latest observation and the latest
prediction ensemble (see Eq.~(\ref{eq:analysis})). We note that it can be preferable to use another
algorithm that is closely related to EnKF, namely the Ensemble Kalman Smoothing method, which finds
optimal correction in light of all past observations~\citep{evensen2000ensemble}.  This method will
be investigated in future work.

\begin{figure}
\begin{center}
\noindent
\includegraphics[width=24pc]{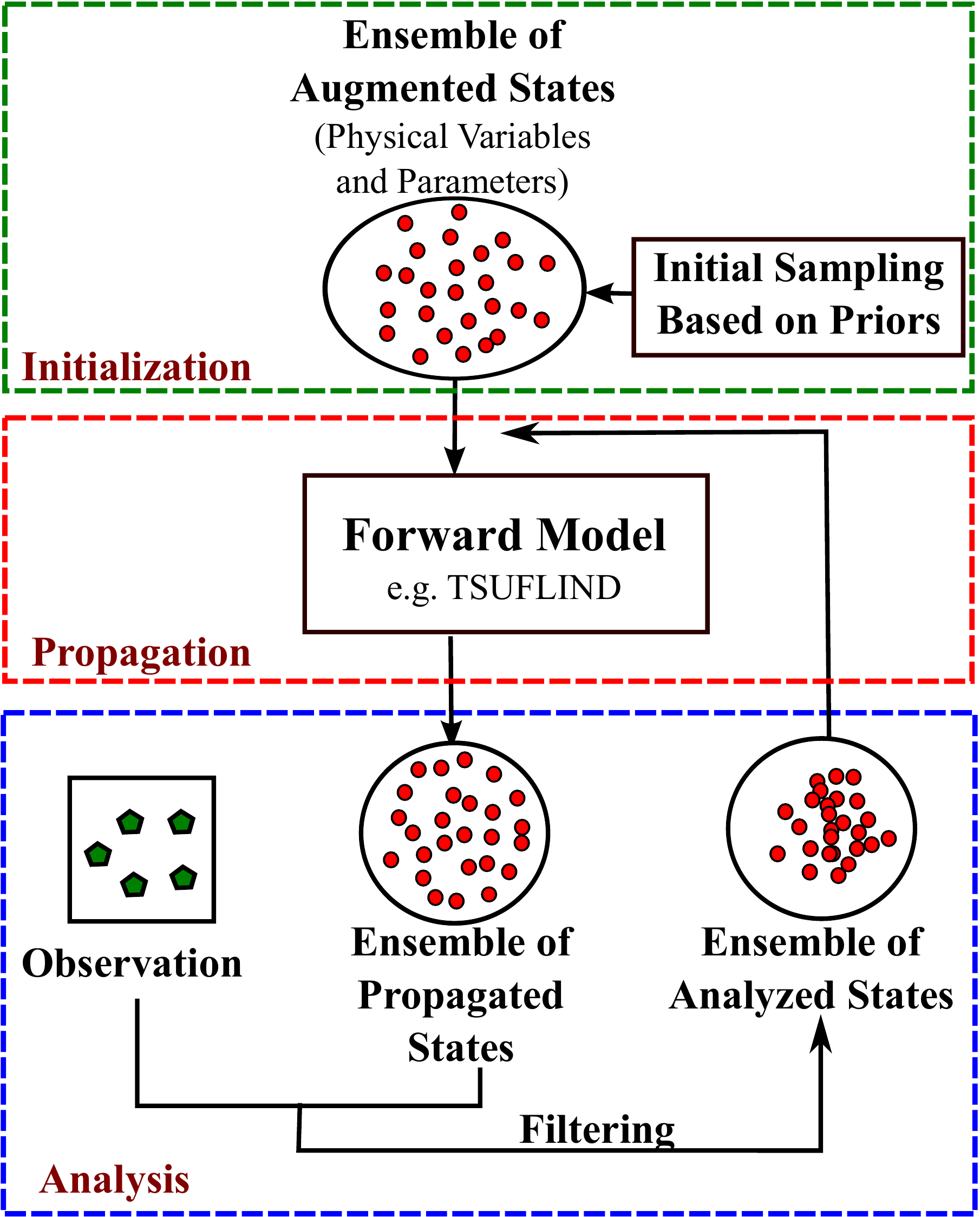}
\caption{Schematic of the inverse modeling approach based on the state augmentation.  The system
  state is augmented to include both the physical state (sediment flux $\zeta_i(t)$) and the
  parameters to be inferred (shear velocity $u_*$ and flow depth $h$). An ensemble representative
  of the augmented state is propagated via the forward model.  The propagated ensemble is then
  updated in the analysis process based on the observation data.  The updated state (physical
  quantities and model parameters) is then propagated to the next time step, and this loop continues
  until all observations are assimilated.}
\label{fig:enkfFlow}
\end{center}
\end{figure}

\section{Computational Setup of Synthetic Cases for Verification}
\label{sec:synthetic}

While EnKF-based Bayesian inferences have been widely used in other communities of geosciences, the
present contribution represents the first attempt in using it for tsunami inversion.  To establish
confidence in the proposed framework for tsunami inversion based on sediment deposits, 
we construct a series of verification cases with synthetic truths to assess the performance of the proposed inversion
scheme. {Furthermore, we test the proposed framework by using a set of field data of the real 
tsunami deposits from the 2006 South Java Tsunami (sections Bunton, sample JTR 6~\citep{Spiske201029}).}

A synthetic case can be generated by running the forward model described in
Section~\ref{sec:met-for} on given a set of tsunami and sediment characteristics (i.e., shear
velocity $\tilde{u}_*$ and flow depth $\tilde{h}$, the range of particle sizes $\tilde{\phi}_i$,
where $\tilde{\cdot}$ indicates synthetic truths). The corresponding tsunami deposits including the
thickness and the grain-size distribution as shown in Fig.~\ref{fig:sed} can thus be obtained.
Subsequently, the sediment flux can be obtained by post-processing the tsunami deposit information
with the procedure explained in Section~\ref{sec:met-inv}. In fact, for the synthetic cases the
sediment fluxes are part of the forward model simulation output, and thus a post-processing
procedure is not required.  Synthetic observations are then generated by adding Gaussian random
noises of standard deviation $\sigma_i$ to the true sediment fluxes, which represent the measurement
and sampling errors in sediment coring operations in the field. The decision to use synthetic cases
instead of realistic cases to verify the proposed method is justified by the fact that the true
tsunami characteristics corresponding to actual field samples are usually unknown, which make them
not ideal for verification purposes. Even if the truth for tsunami characteristics were known, e.g.,
when the flow speed and flow depth were measured from independent sources, it would be difficult to
distinguish the errors due to the forward model inadequacy { and} those due to the inversion
procedure. Therefore, using the synthetic cases allow us to focus on assessing the performance of
the proposed inversion procedure. The merits of the inversion scheme can be assessed by its
capability to reproduce the synthetic truths $\tilde{u}_*$ and $\tilde{h}$, to which the inversion
scheme is blind. {With the established confidence from the verification cases, field data of deposits
from the 2006 Java tsunami event are used as the observations to demonstrate the capability of the
proposed framework in realistic applications.}

For simplicity, we assume that the shear velocity $u_*$ and flow depth $h$, which are to be inferred,
are assumed time-invariant.  In addition, tsunami deposit at only one onshore location is
utilized.  The proposed scheme can be straightforwardly extended to time-varying shear velocity
$u_*(t)$ by incorporating iterations in each time step.  Moreover, for the problems where sediment
deposits are available at multiple locations, the proposed scheme also can be applied by expanding
the state vector to include fluxes at different locations.

{Two verification cases with synthetic observations} 
of increasing difficulty levels are defined. In case~1, the
sediment has a single grain size $\phi = 2.0$, and the only unknown parameter to be inferred is the
shear velocity $u_*$ while the flow depth $h$ is given. In case~2, which is more challenging, the
sediment has a log-normal grain-size distribution with $0 < \phi < 3.25$. Both $u_*$ and $h$ are
unknown and are to be inferred. {The synthetic truths for
$u_*$ and $h$, the prior ensemble means ($\bar{u}_*^{0}$ for both cases and $\bar{h}^{0}$ for case 2
only) are the same for both synthetic cases. For the realistic case with field data (referred to as case 3), 
the mean grain size, largest grain size and smallest grain size are 
$\phi=2.5$, $\phi=0.0$ and $\phi=5.25$, respectively. The mean values of prior ensembles are 
estimated by TSUFLIND~\citep{huitang2015}.
All other parameters, including the forward model time step $\Delta
t$, the data assimilation interval $\Delta N$, and the number of samples $M$ are the same for all 3 
cases.} The parameters are summarized in Table~\ref{Tab:caseSet}.  The
prior ensembles for both parameters are uniformly distributed in the ranges specified in
Table~\ref{Tab:caseSet}, which is representative of the lack of knowledge on the quantities to be
inferred in practical tsunami inversions.  Since the truths of the quantities to be inferred, $\tilde{u}_*$
and $\tilde{h}$, are unknown when performing the inversion, the mean values of prior ensemble $\bar{u}_*^0$
and $\bar{h}^0$ for both quantities are likely to have biases compared to the synthetic truth. This is
reflected in the choice of ensemble mean as shown in Table~\ref{Tab:caseSet}.  The time step $\Delta
t = 0.5$~s is chosen for {all cases}, and the observations of sediment fluxes are assimilated every
$\Delta N = 10$ time steps. The error covariance matrix is chosen as $\mathbf{R} =
\textrm{diag}[\sigma_1^2, \cdots, \sigma_n^2] $ with $\sigma_i$ being the standard deviation of the
white noise added to the true sediment fluxes for each grain size $i$.  The choice of standard
deviation of noises shown in Table~\ref{Tab:caseSet} is based on the error model detailed in a
companion paper.

\begin{table}
\caption{ Physical and computational parameters for test cases.}
\centering
\begin{tabular}{l|c|c|c}
  \hline
  \textbf{Parameters} & \textbf{Case 1}  &  \textbf{Case 2}  & \textbf{Case 3}      \\ 
  \hline
  Inferred parameters &   $u_*$        & $u_*$ and $h$   & $u_*$ and $h$         \\ 
  Grain size $\phi$ $(= - \log_2 [D/D_\textrm{ref}])$ &   $2.0$  &     $0 < \phi < 3.25$   & $0 < \phi < 5.25$ \\
  \hline
  $\Delta t$ & \multicolumn{3}{c}{0.5~s} \\
  Data assimilation interval $\Delta N$ & \multicolumn{3}{c}{10} \\
  Number of samples $M$ & \multicolumn{3}{c}{1000} \\
  std.~$\sigma_i$ of observation error\tablenotemark{a}  & 
  \multicolumn{2}{c}{$\epsilon$ + 0.01$\tilde{\zeta}_i$} & $\epsilon$ + 0.05$\tilde{\zeta}_i$\\ 
  \hline
  Synthetic truth $\tilde{u}_*$ for $u_*$ & \multicolumn{2}{c}{$0.5\ \mathrm{ ms}^{-1}$} & $-$\\
  Range of prior ensemble for $u_*$ \tablenotemark{b} & \multicolumn{2}{c}{0.4 to
    $1.2\ \mathrm{ ms}^{-1}$} & 0.15 to
    $0.45\ \mathrm{ ms}^{-1}$\\
  Mean  $\bar{u}_*^{0}$ of prior ensemble for $u_*$  & \multicolumn{2}{c}{$0.8\ \mathrm{ ms}^{-1}$} & $0.3\ \mathrm{ ms}^{-1}$ \\
  \hline
  Synthetic truth  $\tilde{h}$ for $h$ & 3~m & 3~m & $-$ \\
  Range of prior ensemble for $h$ & $-$ & 2.5 to 7.5~m & 6 to 10~m \\
  Mean   $\bar{h}^0$ of prior ensemble for $h$ &  $-$  &   5~m & 8~m\\
  \hline
\end{tabular}
\tablenotetext{a}{ std.\ denotes standard deviation; the standard deviation of noise added to the
  observation depend on the grain-size class; $\epsilon = 0.125$ $\mathrm{mm^3 s^{-1}}$ is a fixed constant, and
  $\tilde{\zeta}$ is the synthetic truth of the sediment flux, which depends on
  the grain size and vary with time}
\tablenotetext{b}{A smaller range of prior ensemble for $u_*$ from 0.7 to 0.9~$\mathrm{m s^{-1}}$ is also investigated in
  case~1. }
\label{Tab:caseSet}
\end{table}

\section{Results and Discussion}
\label{sec:results}

 \subsection{Case 1: A Single Grain-Size Class and One Unknown Parameter}

 First, a case with simple conditions is studied where all tsunami deposits are in the same
 grain-size class and the shear velocity $u_*$ is the only unknown parameter to be inferred.
 Although this highly simplified scenario is likely to be uncommon in reality, a case with
 controlled conditions is able to provide a reasonable initial assessment of the proposed method.

\begin{figure}[h!]
  \centering
    \includegraphics[width=30pc]{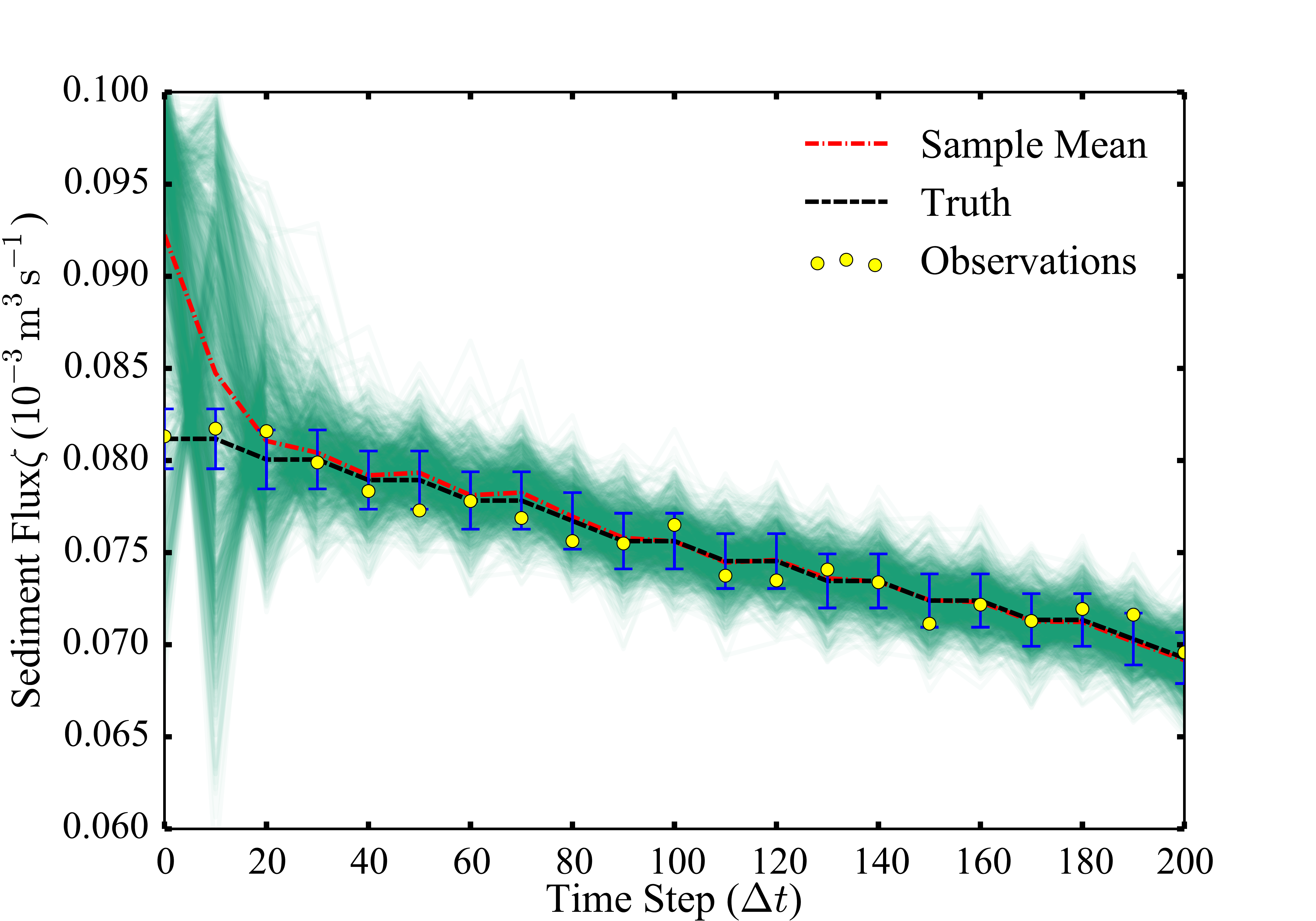}
    \caption{Time series of the sediment fluxes $\zeta(t)$, which is the physical state of the system,
      for a single grain-size class $\phi = 2.0$ during the sedimentation process.  The green (light
      grey) lines show $M=1000$ sediment flux samples, and the yellow (filled) circles indicate the
      observed sediment flux corresponding to the synthetic truth of the shear velocity $\tilde{u}_* =
      0.5\ \mathrm{ms}^{-1}$.  (The 2$\sigma_o$ observation error bar is also plotted in the figure.) 
      The synthetic observation data of sediment fluxes are assimilated to the simulation every
      10 times in the EnKF-based inversion procedure.}
 \label{fig:fluxSingle}s
\end{figure}

 The time series $\zeta (t)$ of the sediment flux, which is the physical state of the system, is
 presented in Fig.~\ref{fig:fluxSingle}. The sediment fluxes for all 1000 samples are shown along
 with the synthetic truth corresponding to the true shear velocity $\tilde{u}_* = 0.5\
 \mathrm{ms}^{-1}$.  It can be seen that for all samples (i.e., prior guesses of shear velocities)
 and the synthetic truth show a general trend that the magnitude of sediment flux decreases with
 time, which is typical for a Rouse sediment concentration profile assumed in the forward model as
 in Eq.~(\ref{eq:ci}).

\begin{figure}[h!]
  \centering
	\includegraphics[width=1.0\textwidth]{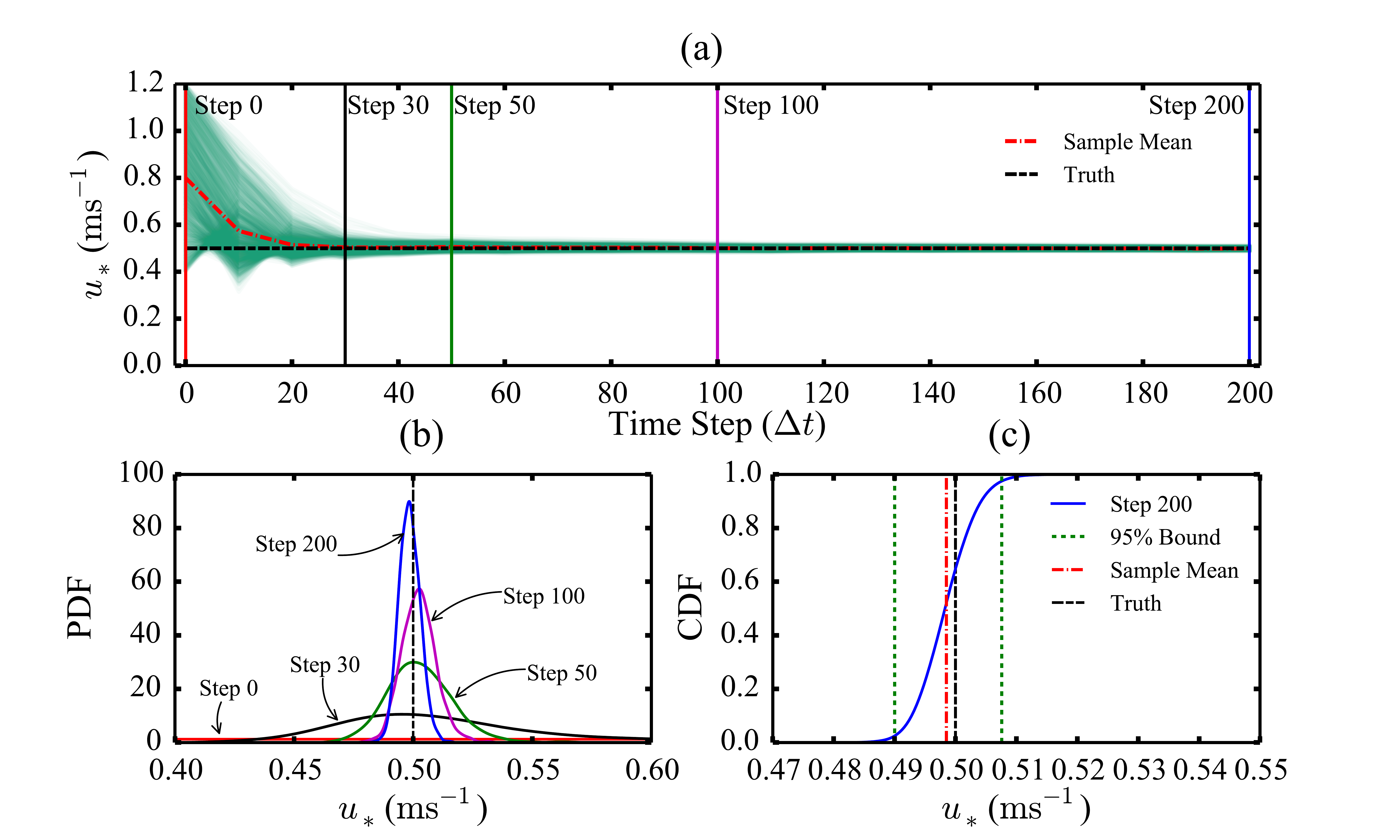}
        \caption{Convergence history of the inferred parameter, the shear velocity $u_*$, for the
          single grain-size case. The plot shows (a) the convergence of the ensemble and the sample
          mean as well as (b) the evolution of the probability density function of the shear
          velocity $u_*$ in the inversion process.  (c)~The corresponding cumulative distribution
          function for the final inference results.} 
        \label{fig:shearSingle}
\end{figure}

The convergence history of the inferred parameter, the shear velocity $u_*$, is shown in
Fig.~\ref{fig:shearSingle}a.  By assimilating the observation data as shown in
Fig.~\ref{fig:fluxSingle}, the shear velocity of all samples and the sample mean gradually converge
to the synthetic truth $\tilde{u}_* = 0.5\ \mathrm{ms}^{-1}$, regardless of their values in the
initial ensemble.  The range of sample scattering for $u_*$, which is $0.8\ \mathrm{ms}^{-1}$ (with
a range from 0.4 to $1.2\ \mathrm{ms}^{-1}$) in the prior ensemble, shrinks to $0.04\
\mathrm{ms}^{-1}$ at the end of the inference. Since the scattering of samples in EnKF-based
inference is indicative of the uncertainty in the inferred results, the reduction of sample
scattering represents the reduction of uncertainties and correspondingly the increase of confidence
in the quantity to be inferred.

The Probability Density Function (PDF) of the shear velocity $u_*$ corresponding to the ensemble at
time steps 0 (initial), 30, 50, 100 and 200 (final) are presented in Fig.~\ref{fig:shearSingle}b,
highlighting the evolution of uncertainty in the inferred quantity.  We can see that the shear
velocity $u_*$ is equally likely between 0.4 and $1.2\ \mathrm{ms}^{-1}$ because of the
non-informative, uniform prior distribution that is chosen.  At time step 50 (after five
observations assimilated) the samples are scattered between approximately 0.47 and $0.54\
\mathrm{ms}^{-1}$.  Moreover, the bias of the ensemble mean compared to the truth has been largely
corrected. As the data assimilation continues, the distribution of $u_*$ continues to shrink and
concentrate towards the truth ($\tilde{u}_* = 0.5\ \mathrm{ms}^{-1}$). The final, posterior
distribution of $u_*$ at time step 200 is Gaussian based on the plot shown in
Fig.~\ref{fig:shearSingle}b. While this could well be the true distribution, it should be
interpreted with caution, since it is well known that the EnKF procedure tend to give a Gaussian
distribution because of its assumptions~\cite{evensen2007data}.  The Cumulative Distribution Function (CDF) of
the posterior distribution is presented in Fig.~\ref{fig:shearSingle}c.  It shows that 95\% of
ensembles have shear velocities $u_*$ between 0.490 and 0.508 $\mathrm{ms}^{-1}$.  In other words, the
inference suggest that there is 95\% probability that the shear velocity $u_*$ falls within the
range above. This credible interval is based on both the prior distribution and the observation
data. 

The uncertainties represented as credible intervals that are obtained in the inversion process is
based on the rigorous Bayesian inference theory, which clearly distinguishes the present method with
existing schemes for tsunami inversion.  The evolution of the inference uncertainty over time as
shown in Fig.~\ref{fig:shearSingle}a and~\ref{fig:shearSingle}b illustrates the Bayesian nature of
the EnKF-based data assimilation method. In the inference, one starts with a subjective prior
distribution on the unknown parameters, which is usually non-informative (see the wide distribution
at time step 0) and is represented with an ensemble. As observation data are assimilated, the
distribution becomes narrower. Meanwhile, the importance of the prior distribution diminishes as
more and more data are assimilated. The remaining uncertainties in the inference results stems from
the uncertainties in the observation data, which are inevitable in field measurement and are
represented with random noise in this study.

In order to demonstrate the robustness of the inversion procedure with respect to the specified prior
distribution, we also tested a prior ensemble (i.e., initial guesses of the parameter to be
inferred) with $u_*$ in the range of 0.7 and $0.9\ \mathrm{m s}^{-1}$. In contrast to the case
presented above, this range does not cover the truth $\tilde{u}_* = 0.5\ \mathrm{m s}^{-1}$. Even
with this overly confident prior distribution, almost identical inversion results were obtained. In
fact, the differences between the convergence history of $u_*$ disappear after a few assimilation
steps. As such, detailed results for this case are omitted here for brevity.

\begin{figure}[h!]
  \centering
	\includegraphics[width=0.6\textwidth]{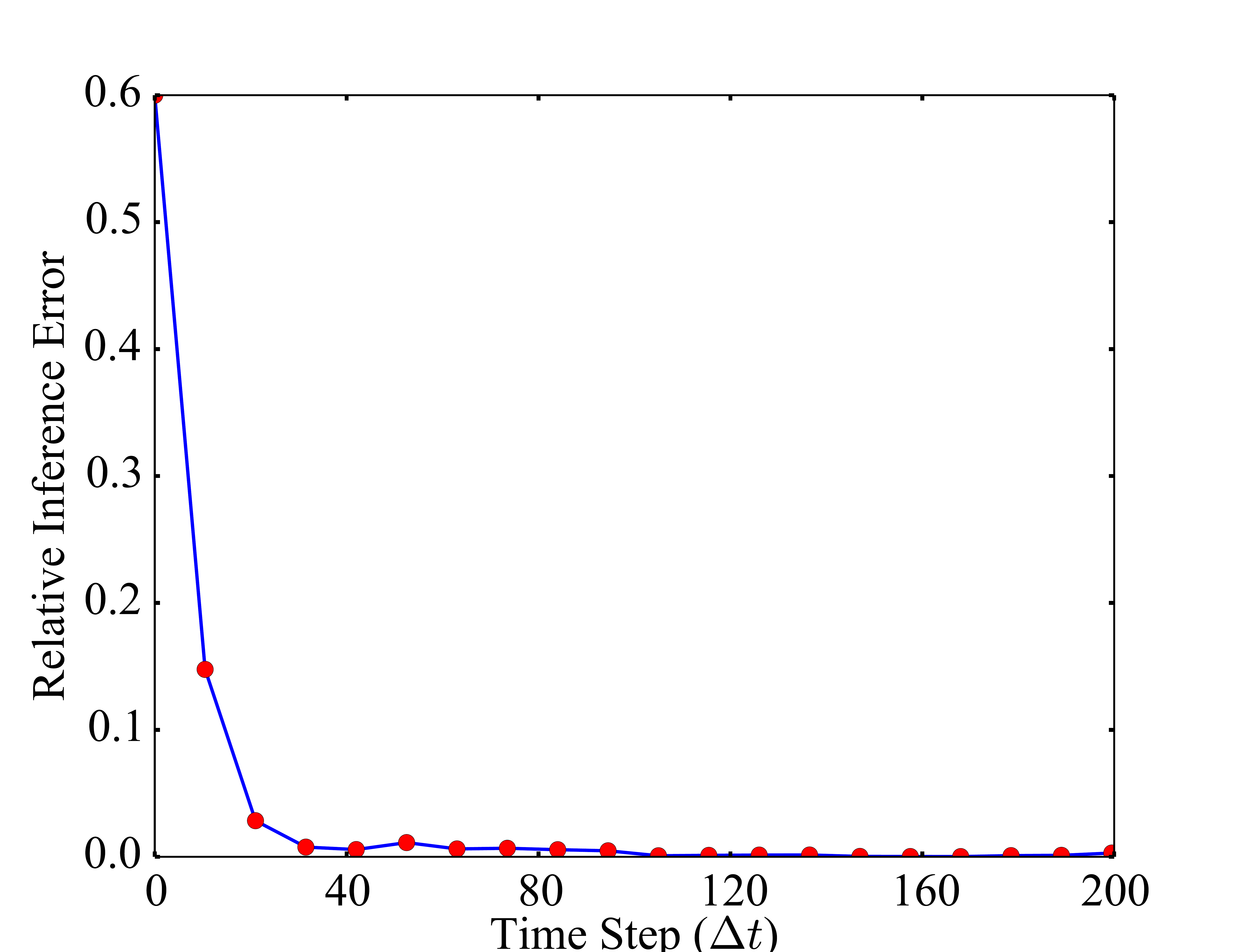}
        \caption{The $L_2$ norm of the inference error for
 the shear velocity, defined as the difference $| (\bar{u}_* - \tilde{u}_*) / \tilde{u}_* |$ between the
 ensemble mean $\bar{u}_*$ and the synthetic truth $\tilde{u}_*$.} 
        \label{fig:shearErr}
\end{figure}

The relative inference error for the shear velocity is presented in Fig.~\ref{fig:shearErr}, which
is defined as the $L_2$ norm $| (\bar{u}_* - \tilde{u}_*) / \tilde{u}_*|$ of the difference between
the ensemble mean $\bar{u}_*$ and the synthetic truth $\tilde{u}_*$.  It can be seen that the
inference error decreases dramatically within the first few data assimilations steps, from $0.6$ for
the initial prior ensemble (time step 0) to $0.025$ after five observations are assimilated (at time
step 50). This finding is consistent with the decrease of the ensemble scattering (indicating
inference uncertainties) as shown in Fig.~\ref{fig:shearSingle}a.

\subsection{Case 2: Multiple Grain-Size Classes and Two Unknown Parameters} 

\begin{figure}[h!]
  \centering
    \includegraphics[width=30pc]{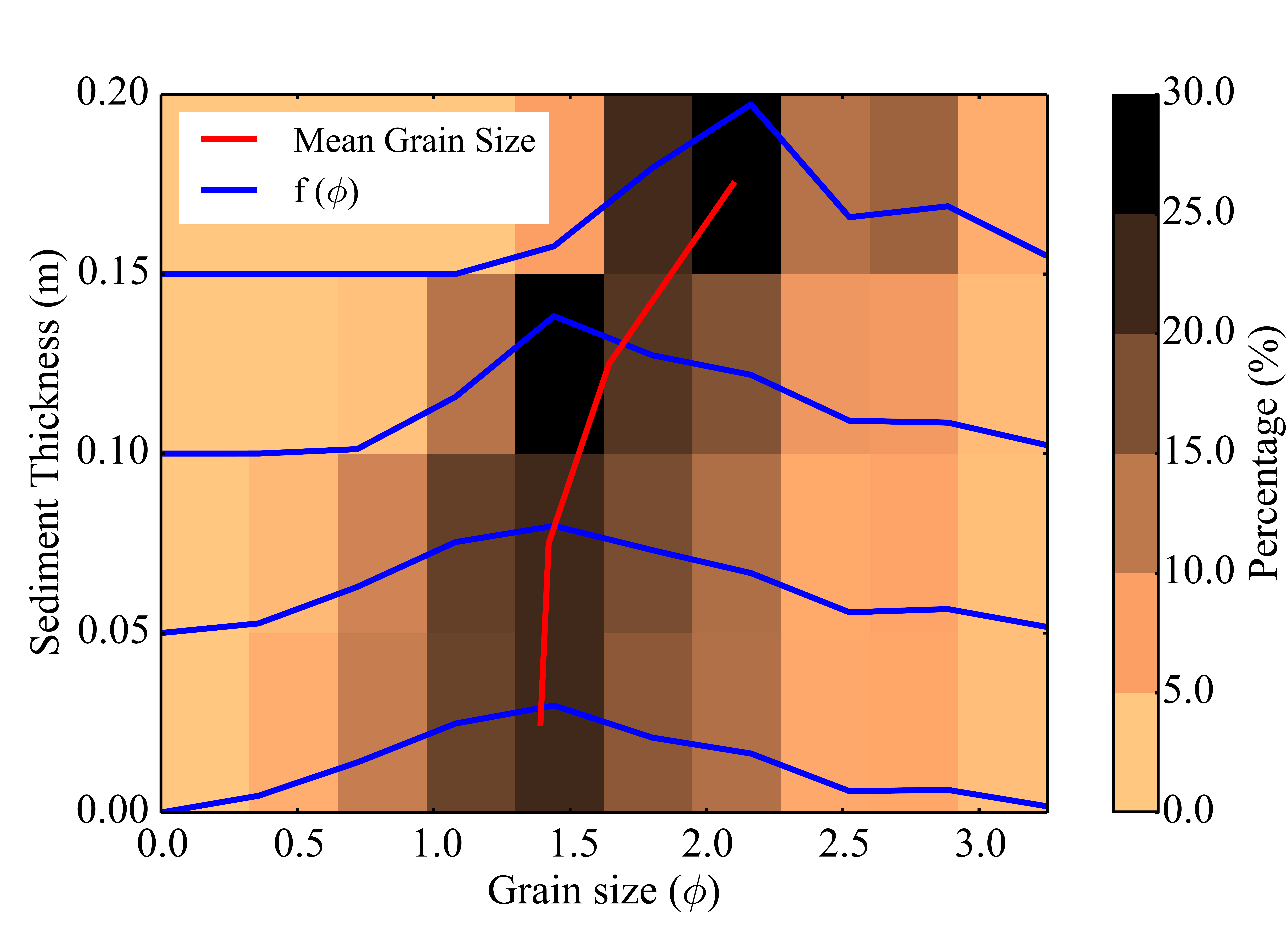}
  \caption{Hypothetical 0.2 m thick tsunami deposit for verification cases: vertical grading in grain-size distributions
  (blue line) and mean grain-size (red line) for sediment intervals. The grain-size distribution of
  the entire tsunami deposit is a log-norm distribution.} 
\label{fig:sed}
\end{figure}

A tsunami inversion problem with wide grain-size distribution is studied to demonstrate the
capability of the proposed method in a more realistic scenario. Both the shear velocity $u_*$ and
the flow depth $h$ are unknown and the two parameters must be inferred simultaneously. The input to
the tsunami inversion procedure is the analyzed results of tsunami deposit column as shown in
Fig.~\ref{fig:sed}.  The grain-size distribution along the depth of the sediment column is obtained
from a forward simulation with shear velocity and flow depth as specified in
Table~\ref{Tab:caseSet}. The particle sizes range from 0 to 3.25 in $\phi$ scale, or equivalently
from 1~mm to 0.105~mm.  The range is divided into 10 grain-size classes, $\phi_i$ with $i = 1,
\cdots, 10$, equally spaced in $\phi$ scale. The superscripts are indices of the grain-size classes.

\begin{figure}[h!]
  \centering
	\includegraphics[width=1\textwidth]{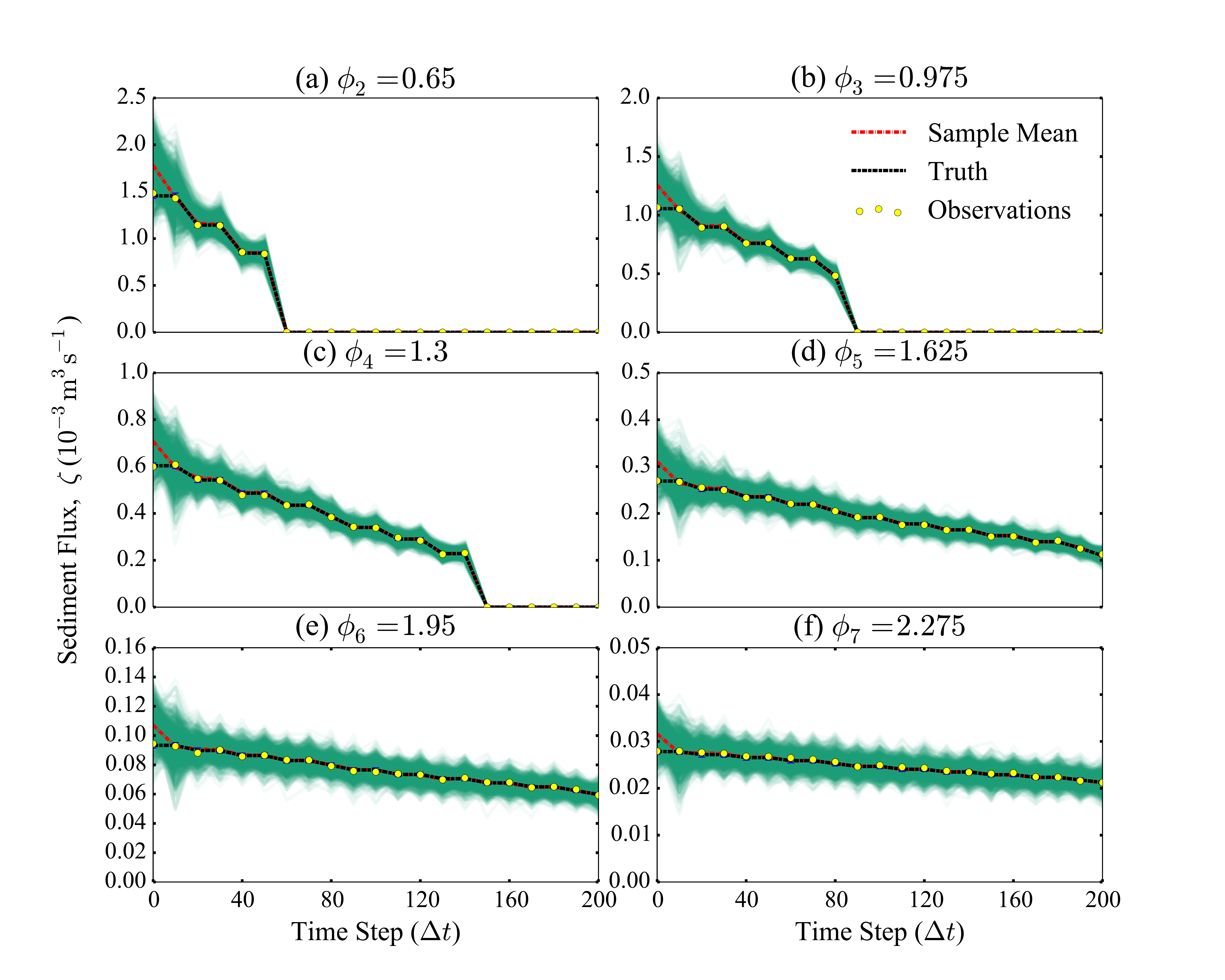}
        \caption{Time series of sediment fluxes $\zeta(t)$ for six (among a total of 10) grain-size
          classes during the sedimentation: (a)~$\phi_2 = 0.65$, (b)~$\phi_3 = 0.975$, (c)~$\phi_4 =
          1.3$, (d)~$\phi_{5} = 1.625$, (e)~$\phi_{6} =1.95$, and (f)~$\phi_{7} = 2.275$.  The
          green (light grey) lines show $M=1000$ samples of sediment flux time series, and the yellow
          (filled) circles indicate the observed sediment flux corresponding to the synthetic truth,
          i.e., shear velocity $\tilde{u}_* = 0.5\ \mathrm{m s}^{-1}$ and flow depth $h = 3$~m.
          synthetic observation data of sediment fluxes are assimilated
          to the simulation every 10 times in the EnKF-based inversion procedure. }
 \label{fig:fluxMulti}
\end{figure}

The time series of sediment fluxes for six representative grain-size classes are shown in
Fig.~\ref{fig:fluxMulti}, i.e., $\phi_2 = 0.65$, $\phi_3 = 0.975$, $\phi_4 = 1.3$, $\phi_{5} =
1.625$, $\phi_{6} = 1.95$, and $\phi_{7} = 2.275$.  As in the case with a single grain-size class,
the sediment fluxes $\zeta_i(t)$ for all grain-size classes decrease as the deposition proceeds, and
the uncertainties represented by the ensemble scattering are reduced as the observations are
assimilated. Moreover, the sediment fluxes for the coarse grains (corresponding to smaller $\phi$
values, e.g., $\phi_2 = 0.65$, $D = 0.637$~mm as shown in Fig.~\ref{fig:fluxMulti}a) decrease more
rapidly than those for the finer grains with larger $\phi$ values.  It can be seen from
Fig.~\ref{fig:fluxMulti}a that the coarsest ($\phi_2 = 0.65$) among the six grain-size classes
completed sedimentation in 60 time steps (i.e., 30 seconds since $\Delta t = 0.5$~s). In contrast,
the finest grain-size class ($\phi_{7} = 2.275$, $D = 0.105$~mm) among the six has a more uniform
sediment flux throughout the entire sedimentation process.  Again, this is attributed to the assumed
Rouse profile. It is also noted that the relative uncertainties (ensemble scattering) in the
sediment fluxes at the end of the inversion are larger for the fine grains than for the coarse
grains. This is due to the random noises added to the true sediment flux when generating synthetic
observations. As shown in Table~\ref{Tab:caseSet}, the standard deviation $\sigma_i$ of the noise
has a fixed component ($\epsilon$) and a component $0.01 \tilde{\eta}_i$ proportional to the truth.
Since the sediment flux for the finer grains are smaller in absolute value, as can be seen from the
different vertical axis ranges in panels (a)--(f), the relative observation uncertainties are thus
larger for the sediment fluxes of the finer grain sizes. Consequently, the uncertainties in ensemble
directly reflect the uncertainties in the observations.

\begin{figure}[h!]
  \centering
    \includegraphics[width=40pc]{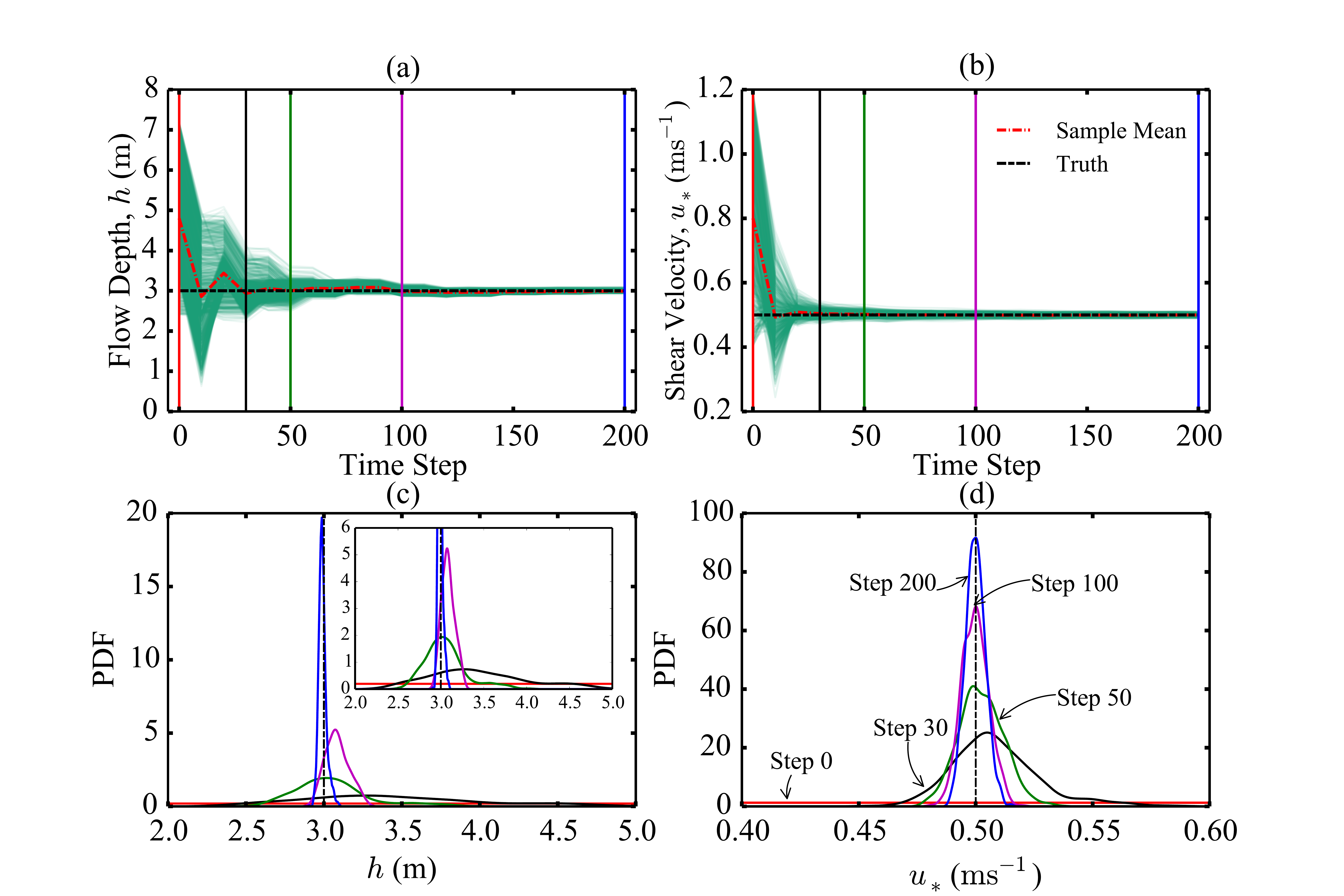}
    \caption{Convergence history of the inferred parameters, the shear velocity $u_*$ and the flow
      depth $h$.  Panels (a) and (b) show the evolution of the ensemble and the sample mean for the
      shear velocity and the flow depth, respectively.  Panels (c) and (d) show the evolution of
      probability density functions for the shear velocity and the flow depth, respectively, during
      the inversion process. In panel (c) a zoomed-in view is presented as  inset to show the
        detailed results in the first few steps.}
    \label{fig:shearTwo}
\end{figure}

The convergence history of the two parameters to be inferred, shear velocity and flow depth, are
presented in Fig.~\ref{fig:shearTwo}. It can be seen from Fig.~\ref{fig:shearTwo}a
and~\ref{fig:shearTwo}b that the sample means for both parameters (red dotted lines) converge to the
synthetic truths (horizontal black dashed lines) within 50 time steps, i.e., after five observations
are assimilated, although the prior ensemble means deviate significantly from the truths at time
step 0.  Moreover, the uncertainties as indicated by the ensemble scattering are reduced
significantly. We note two points here. First, comparison between Fig.~\ref{fig:shearTwo}a
and~\ref{fig:shearTwo}b suggests that the shear velocity converges to the truth faster than the
flow depth does. Since EnKF inversion procedure depends on correlation to make inferences, this
seems to indicate that the sediment fluxes, which are the observed physical state, are more
sensitive to the shear velocity than that to the flow depth. Second, the convergence of shear
velocity in the multiple grain-size case as shown in Fig.~\ref{fig:shearTwo}b is slightly faster
than that in the single grain-size class case in Fig.~\ref{fig:shearSingle}a. A major difference
between the two cases is that the state in the multiple grain-size case is the sediment flux for all
grain-size classes, i.e., a vector of size 10 with one component corresponding to each grain-size
class.  Accordingly, an observation in the multiple grain-size case is also a vector of size
10. This is much more information than in the single grain-size class case, where the state and
observations are only scalars. More information in the observation leads to more accurate
inference. Intuitively, one would expect the multiple grain-size class case to be more difficult,
particularly considering the fact that there are two unknown parameters. Indeed, this apparently
counterintuitive finding suggests that the setup here is not entirely realistic in that we used
similar levels of relative error for both the single and multiple grain-size cases. In the field,
when a sediment core of a given size is divided to yield grain-size distributions for multiple
grain-size classes, the relative measurement error in the obtained grain-size distribution would
inevitably increase compared to a single grain-size class case.  Therefore, it would have been more
realistic to use a larger observation error for the multiple grain-size case. The effect of
observation errors on the inference results and the optimization of number of grain-size classes
from a given sediment core are topics of future work.  The decrease of uncertainties in the inferred
parameters can be clearly seen in the probability density functions shown in
Fig.~\ref{fig:shearTwo}c and ~\ref{fig:shearTwo}d.  For both parameters, the probability density
functions shrink continuously towards the truth, indicating the gain of confidence as more
observations are assimilated.  Overall, in this more realistic and challenging test case, we found
that the proposed method has equally satisfactory performance compared to that in the single
grain-size case.

{ 

\subsection{Case 3: Realistic Application} 
For this case, a deposit column with a 0.12~m thickness from the 2006 South Java Tsunami is used as the observation. 
We simultaneously infer both unknown parameters, shear velocity $u_*$ and flow depth $h$. 
The particle sizes range from 0 to 5.25 in $\phi$ scale, or equivalently from 1~mm to 0.0263~mm in diameter. 
We divided this range into 15 grain-size classes $\phi_i$ with $i = 1, \cdots, 15$, equally spaced in 
$\phi$ scale. 

\begin{figure}[h!]
  \centering
    \includegraphics[width=40pc]{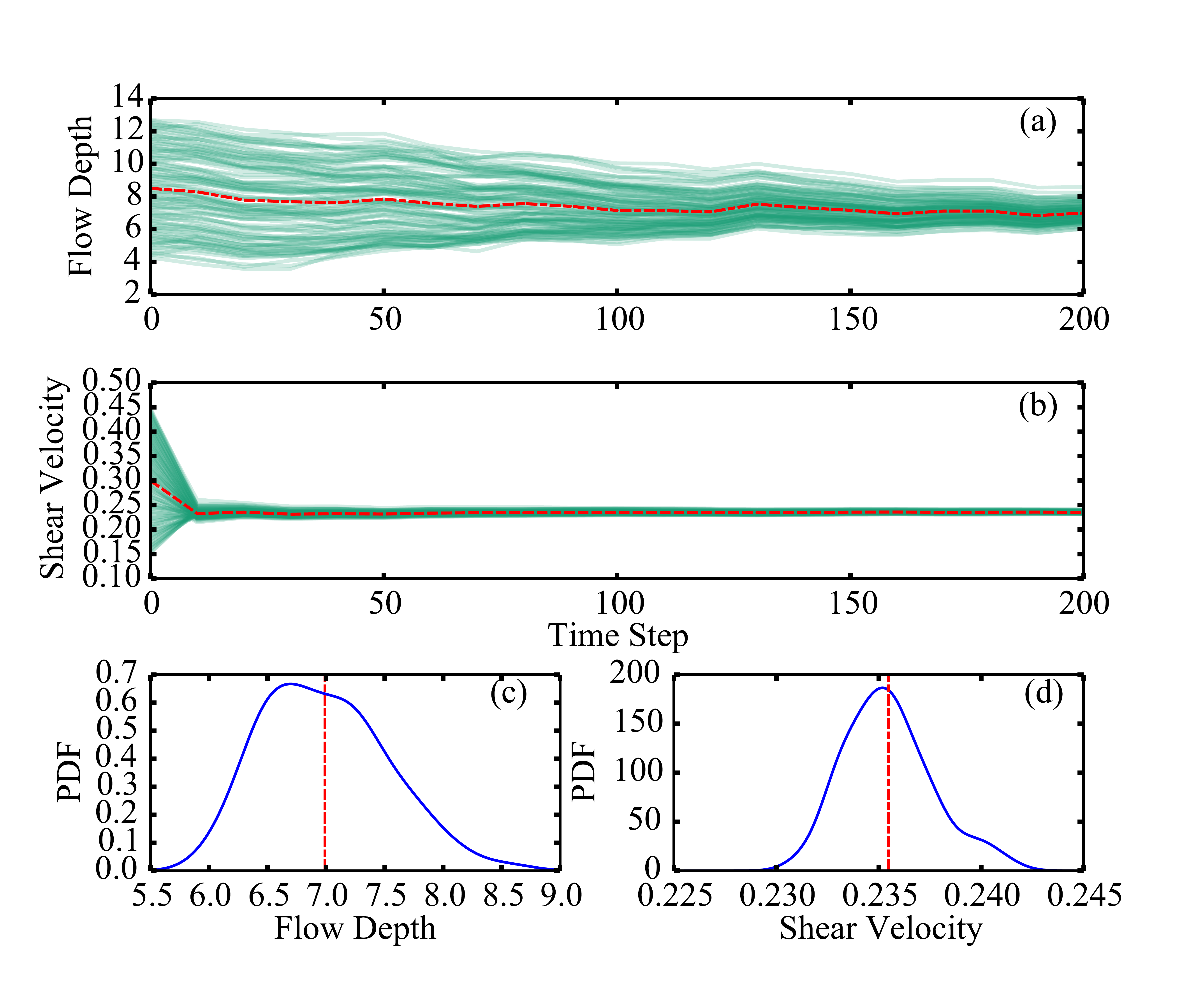}
    \caption{Convergence history of the inferred parameters, the shear velocity $u_*$ and the flow
      depth $h$, for case 3.  Panels (a) and (b) show the evolution of the ensemble and the sample mean for the
      shear velocity and the flow depth, respectively.  The green (light grey) lines denote samples and the red
      dashed lines denote sample means. Panels (c) and (d) show the probability density functions
      for the inferred shear velocity and the flow depth, respectively, at the final time step (200).
      The sample mean is denoted with red dashed line.}
    \label{fig:shearReal}
\end{figure}

The convergence history of shear velocity and flow depth is presented in Fig.~\ref{fig:shearReal}.  
We can see that the scattering of both parameters is reduced.  Within 200 steps,  
flow depths in most of the samples converges to the range of 6 to 8~m from initial range of 4 to 12~m.
Meanwhile, shear velocities in the samples converges to a small interval around 0.235 $\mathrm{ms^{-1}}$. 
Similar to what has been shown in synthetic case 2, the shear velocity converges faster than the flow depth
does. The scattering of shear velocity samples is significantly reduced after 200 steps,  while the flow depth
samples are still largely scattered, indicating a relatively large posterior inference uncertainty. This is because
the sediment fluxes are not sensitive to the flow depth, especially when the flow depth is large. 
Specifically, the Rouse profiles depicted in Fig.~\ref{fig:forward} shows that the sediment concentration is
low in the upper region of the water column.  When the flow is deep, the change of its depth does not significantly affects
the sediment concentration in the upper layers of the water, and thus the sediment fluxes are not sensitive to the flow depth. 
The probability density functions of the flow depth and shear velocity in the last time step (posterior)
are shown in Fig.~\ref{fig:shearReal}c and~\ref{fig:shearReal}d, respectively. The posterior distributions of
flow depth and shear velocity are approximately Gaussian with mean values of 7.0~m and 0.236~ms$^{-1}$,
respectively.

\begin{figure}[h!]
  \centering
    \includegraphics[width=40pc]{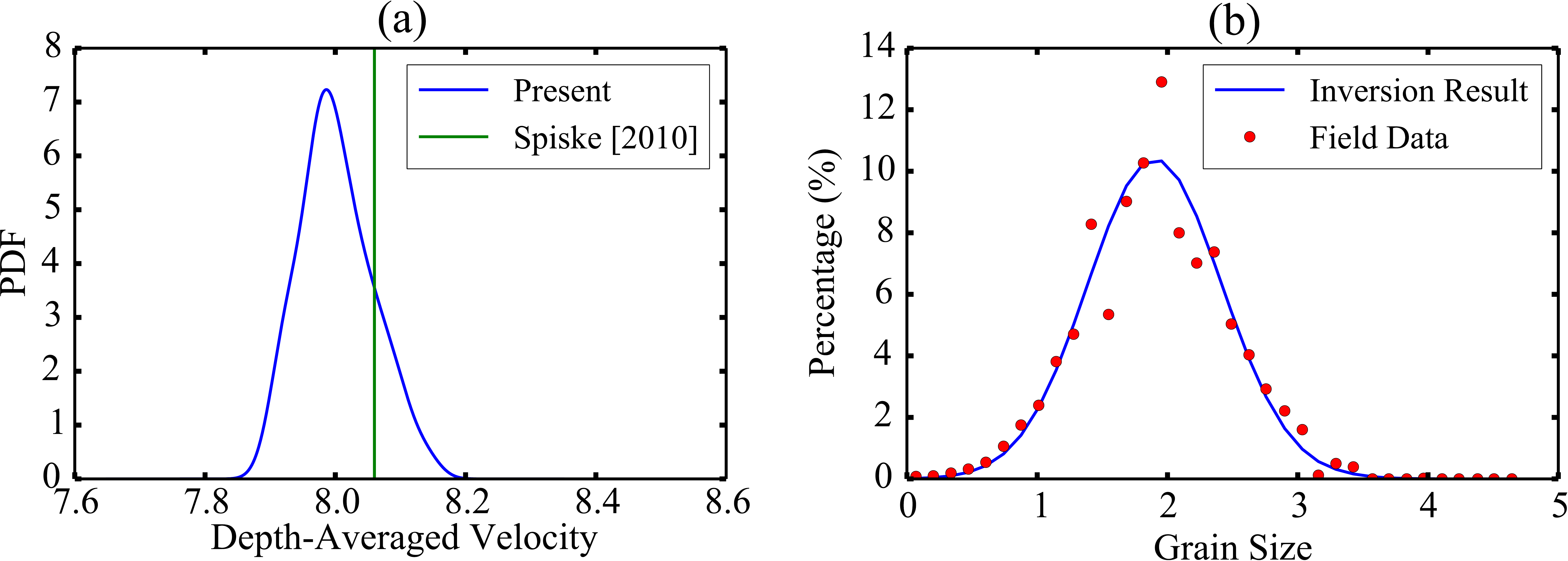}
    \caption{Comparison of two inferred QoIs, the depth-averaged velocity and the grain-size distribution, with
    the previous inference of ~\citep{Spiske201029} (depth-averaged velocity) and field data (grain-size distribution), 
    respectively. Panel (a) shows the comparison of depth-averaged velocity calculated based on the inferred 
    shear velocity $u_*$ and water depth $h$ from the proposed inversion scheme with that of 
    TsuSedMod results~\citep{Spiske201029}. Panel (b) compares the grain-size distribution obtained from 
    the forward simulation with the inferred parameter $u_*$ and $h$ with the field data.  
     }
    \label{fig:shearCompare}
\end{figure}

Since there are no ground truths of flow depth and shear velocity available for validation in this realistic case, we
have to validate the inference results indirectly. Specifically, the depth-averaged velocity is computed by using
the inferred shear velocity and flow depth based on Eq.~\ref{eq:uz} and is compared
with that obtained by \citet{Spiske201029} with the inversion model TsuSedMod. The 
comparison is shown in Fig.~\ref{fig:shearCompare}a. We can see that the depth-averaged velocity
from our inference is distributed from 7.8 to 8.2~ms$^{-1}$ with a mean value of 8.0~ms$^{-1}$. 
This posterior credible interval covers the inference result (8.06~ms$^{-1}$) of \citet{Spiske201029}, although the
sample mean from our inference is slightly smaller. A possible explanation of the discrepancy is that 
\citet{Spiske201029} used the entire deposit for the TsuSedMod model to perform
the inversion, but the sediments in the bottom layers of the deposit may be transported by bed load.  
Since the TsuSedMod model assumes the sediments is formed only by suspension load~\citep{jaffe2007simple}, 
the velocity obtained by \citet{Spiske201029} may have been overestimated.
In order to compare the inversion results with the field data, we perform forward simulations of sedimentation
with the inferred parameters, i.e., mean values of posterior flow depth and shear velocity. 
Figure~\ref{fig:shearCompare}b compares the grain-size distribution obtained based on the 
inferred parameters with that from the field data. It can be seen that the inferred grain-size distribution 
agrees very well with the field data, which shows a satisfactory performance of the proposed 
inversion scheme in the realistic case. However, it should be noted that this indirect
validation of the inference does not guarantee the inferred quantities (i.e., flow depth and shear velocity) are
accurate, since the model error of the forward model affects the accuracy of the inversion.
If the forward model adequately describes the physics of the sedimentation process, 
the inference results are accurate. Therefore, we can improve the forward model to achieve more accurate inference results.  
}

\section{Conclusion}
In this work, we proposed a novel inversion scheme based on Ensemble Kalman Filtering to infer
tsunami flow speed and flow depth from tsunami deposits.  In contrast to traditional data
assimilation methods using EnKF, an important novelty of the current work is that the system state
is augmented to include both the physical variables (sediment fluxes) and the unknown parameters to
be inferred, i.e., shear velocity and flow depth.  Consequently, the unknown tsunami
  characteristics are inferred in a rigorous Bayesian way with quantified uncertainties, which
  clearly distinguishes our method with existing tsunami inversion schemes. Two test cases with
synthetic observation data are used to verify the proposed inversion scheme.  Numerical results show
that the tsunami characteristics inferred from the tsunami deposit information have favorable
agreement with the synthetic truths, which demonstrated the merits of the proposed tsunami inversion
scheme. {A realistic case with field data is studied, and the results are compared to those
obtained with a previous inversion model TsuSedMod and are validated by the field data. The 
comparisons indicate a satisfactory performance of the proposed inversion scheme on realistic applications.}
The proposed inversion scheme is a promising tool for the study of paleo tsunamis in the
interrogation of sediment records to infer tsunami characteristics.

\appendix

\section{Detailed Algorithm for Ensemble Kalman Filtering}
\label{app:enkf}

The algorithm of the ensemble Kalman filtering for data assimilation and inverse modeling is
summarized below. 

Given the prior distribution of the parameters to be inferred (shear velocity $u_*$ and flow depth
$h$) and sediment flux observations with error covariance matrix $\mathbf{R}$, the following steps
are performed:
\begin{enumerate}
\item \textbf{(Sampling step)} Generate initial ensemble $\{{\mathbf{x}_j}\}_{j = 1}^M$ of size $M$,
  where the augmented system state is:
  \begin{equation}
    \label{eq:ini-x}
    \mathbf{x} = [\zeta_1, \cdots, \zeta_n,  u_*, h]'.
    \notag
  \end{equation}

\item \textbf{(Prediction step)} 
  \begin{enumerate}
  \item Propagate the state from current state at time $t$ to the next assimilation step $t + \Delta
    T$ with the forward model TSUFLIND, indicated as $\mathcal{F}$,
  \begin{equation}
    \label{eq:forward}
    \hat{\mathbf{x}}_j^{(t+\Delta T)} = \mathcal{F} [ \mathbf{x}_j^{(t)} ]
    \notag
  \end{equation}
  in which $\Delta T = \Delta N \Delta t$, indicating that the observation data is assimilated every
  $\Delta N$ time steps.

\item
  Estimate the
  mean $\bar{\mathbf{x}}$ and covariance $\mathbf{P}^{(n+1)}$ of the ensemble as:
  \begin{subequations}
    \begin{align}
      \bar{\mathbf{x}}^{(t+\Delta T)} = \frac{1}{M}\sum_{j=1}^N{\hat{\mathbf{x}}^{(t+\Delta T)}_j}    \label{eq:mean} \\
      \mathbf{P}^{(t+\Delta T)} = \frac{1}{M-1} \sum_{j = 1}^{N} {\left( \hat{\mathbf{x}}_j\hat{\mathbf{x}}_j' -  
          \bar{\mathbf{x}}\bar{\mathbf{x}}' \right)^{(t+\Delta T)}}
 \label{eq:cov} 
    \end{align}
  \end{subequations}
\end{enumerate}

\item  \textbf{(Analysis step)}
  \begin{enumerate}
  \item Compute the Kalman gain matrix as:
    \begin{equation}
      \label{eq:kalman-gain}
      \mathbf{K}^{(t+\Delta T)} = \mathbf{P}^{(t+\Delta T)} \mathbf{H}' (\mathbf{H} \mathbf{P}^{(t+\Delta T)} \mathbf{H}' + \mathbf{R})^{-1} 
      \notag
    \end{equation}

  \item Update each sample in the predicted ensemble as follows:
        \begin{equation}
      \label{eq:update}
      \mathbf{x}_j^{(t+\Delta T)} = \hat{\mathbf{x}}_j^{(t+\Delta T)} + \mathbf{K}
      (\boldsymbol{\zeta}_j  - \mathbf{H} \hat{\mathbf{x}}_j^{(t+\Delta T)}) 
      \notag
    \end{equation}
  \end{enumerate}

\item Repeat the prediction and analysis steps until  all observations are assimilated.
\end{enumerate}

\section{Notation}

\begin{tabbing}
  0000000\= this is definition\kill 
  $h$ \> water depth \\ 
  $n$ \> number of grain-size classes \\
  $U$ \> depth-averaged flow speed\\
  $u_*$ \> shear velocity \\
  $w$ \> settling velocity of particles  \\
  $z$ \> elevation above the bed   \\
  $z_0$ \>  total roughness of the bed   \\
  $\mathbf{x}$ \> augmented system state   \\
  $C(z)$ \> vertical  profile of sediment concentration \\
  $C_{i, 0}$ \> sediment concentration of class $i$ at the bed  \\
  $C_{0}$  \> total sediment concentration at the bed   \\
  $D$ \> mean particle diameter  \\  
  $\mathcal{F}$ \> forward model for sedimentation \\
  $K$ \> eddy viscosity \\
  $M$ \> number of samples \\
  $N$ \> number of time steps \\  
  $T$ \> total time for all sediments to settle \\
  $U$ \> depth-averaged flow velocity  \\
  $\mathbf{R}$ \> observation covariance matrix  \\
  $\mathbb{R}^n$ \>  space of $n$-dimensional real-valued vectors  \\
  $\mathbf{P}$ \> ensemble covariance matrix  \\
  $\mathbf{K}$ \> Kalman gain matrix \\
  $\mathbf{H}$ \> observation matrix  \\

{Greek letters}{}\\
  $\zeta$ \> sedimentation flux \\
  $\kappa$ \> Von Karman constant \\
  $\phi$ \> logarithmic scale of particle diameter \\
  $\sigma_i$ \> standard deviation of observation noise for the $i^{th}$ grain-size class  \\
  $\Delta t$ \> time step\\
  $\Delta z$ \> water column layer thickness \\
  $\Delta \eta$ \> deposit layer thickness \\    
  $\Delta N$ \>  time steps between two observations\\ 
  $\Delta T$ \>  time interval between two observations\\ 

{Subscripts/Superscripts}{}\\
 $i$ \> index of grain-size class \\
 $j$ \> index of ensemble member \\
 $l$ \> indices of time step, sediment layer, and water column layer \\
 $0$ \> initial ensemble  \\

{Decorative symbols}{}\\
$\tilde{\Box}$ \> synthetic truth \\
$\bar{\Box}$ \> mean \\
$\hat{\Box}$ \> forecast state in EnKF  \\
$\Box '$ \> vector/matrix transpose \\
\end{tabbing}

\begin{acknowledgments}
  JXW and HX gratefully acknowledge partial funding of graduate research assistantship from the
  Institute for Critical Technology and Applied Science (ICTAS, grant number 175258) in this effort.
\end{acknowledgments}



%
%
\end{article}
\end{document}